\documentclass[fleqn,10pt]{wlscirep}
\usepackage[utf8]{inputenc}
\usepackage[T1]{fontenc}
\usepackage{graphicx}
\pdfoutput=1
\title{An Exploration of Multimodality and Data Augmentation for Dementia Classification}

\author[1,2,*]{Kaiying Lin}
\author[1]{Peter Y. Washington}

\affil[1]{University of Hawai'i, Department of Information and Computer Science, Honolulu, 96822, USA}
\affil[2]{University of Hawai'i, Department of Linguistics, Honolulu, 96822, USA}

\affil[*]{kylin@hawaii.edu}

\begin{abstract}
Dementia is a progressive neurological disorder that profoundly affects the daily lives of older adults, impairing abilities such as verbal communication and cognitive function. Early diagnosis is essential for enhancing both lifespan and quality of life for affected individuals. Despite its importance, diagnosing dementia is complex and often necessitates a multimodal approach incorporating diverse clinical data types. In this study, we fine-tune Wav2vec and Word2vec baseline models using two distinct data types: audio recordings and text transcripts. We experiment with four conditions: original datasets versus datasets purged of short sentences, each with and without data augmentation. Our results indicate that synonym-based text data augmentation generally enhances model performance, underscoring the importance of data volume for achieving generalizable performance. Additionally, models trained on text data frequently excel and can further improve the performance of other modalities when combined. Audio and timestamp data sometimes offer marginal improvements. We provide a qualitative error analysis of the sentence archetypes that tend to be misclassified under each condition, providing insights into the effects of altering data modality and augmentation decisions. 
\end{abstract}

\begin{document}

\flushbottom
\maketitle
%
%
\thispagestyle{empty}

\section*{Introduction}

Dementia is a complex syndrome characterized by a decline in cognitive functions such as memory, thinking, and reasoning. An estimated 47.5 million people worldwide are affected by dementia, with some experiencing severe emotional and language impairments. Recognizing these serious consequences underscores the critical importance of early diagnosis in clinical practice.

The diagnostic process for dementia often includes a comprehensive review of the patient's medical history, genetic testing, psychiatric evaluations, and cognitive assessments, often supplemented by neuroimaging techniques. Given the multi-faceted nature of dementia diagnosis, there is growing interest in streamlining the process through scalable, accessible, and low-cost methods. Among the cognitive deteriorations caused by dementia, verbal and speech impairments are notable, making verbal fluency a promising early diagnostic indicator.

One widely-used assessment tool for verbal fluency is the Verbal Fluency Test (VFT), which measures both the speed and thematic organization of word production. The Dementia Databank \cite{Lanzi2023}, the largest publicly available dataset on dementia, offers data derived from patients undergoing these assessments. These datasets, consisting of audio recordings and text transcripts, provide a rich resource for machine learning (ML) models aimed at distinguishing between healthy individuals and those with dementia.

Previous research efforts have employed machine learning models for dementia detection, with some studies fine-tuning pre-existing language models \cite{Yuan2020DisfluenciesAF}, and others developing models from the ground up \cite{luz2020alzheimers}. However, most prior work has focused on singular data types—either audio  \cite{torre2021, Chlasta2021} or text data \cite{Guo2021}—for model training. Fewer studies have explored the synergistic effects of integrating these different data types\cite{sarawgi2020multimodal, hlédiková2022data}.

In this study, we integrate multiple data modalities—including audio, text, and timestamps—sourced from the Dementia Databank to offer a holistic approach for dementia detection. Leveraging pre-trained embeddings like Wav2vec and Word2vec, we strive to boost both the efficiency and effectiveness of our diagnostic models. Our findings indicate that text-based models, particularly when augmented with additional data, deliver robust outcomes and significantly enhance performance in multimodal settings. In contrast, audio and timestamp data contribute only modest improvements. The remainder of this paper is organized as follows: Section 2 offers an overview of relevant literature; Section 3 details our six experimental models, each employing different combinations of data modalities and data augmentation techniques; Section 4 outlines our experimental methodology and results; Section 5 discusses provides concluding observations and future direction. To our knowledge, this is the first study to incorporate time embeddings in conjunction with text and audio data for a comprehensive, multimodal approach to dementia diagnosis.

\section*{Related work}

Many previous research efforts have focused primarily on detecting a specific type of dementia, such as Alzheimer's Disease (AD). Within the Dementia Databank, the Alzheimer's Dementia Recognition through Spontaneous Speech (ADReSS) Challenge \cite{luz2020alzheimers} offers multiple shared tasks, allowing researchers to base their methodologies on common datasets for comparative analysis. Prior AD detection techniques in these shared tasks have employed fine-tuning of existing models, data augmentation, and feature engineering. Studies that utilize feature engineering \cite{luz2020alzheimers, Balagopalan, sarawgi2020multimodal, Chlasta2021} have extracted audio and text features—either manually or through existing packages—and trained models on binary classification tasks.

Other research efforts have fine-tuned pre-trained language models like BERT \cite{devlin2019bert} to achieve similar goals \cite{Balagopalan, Guo2021}. Data augmentation strategies have also been applied to mitigate the challenge of limited data availability \cite{hlédiková2022data}.

In addition to aiming for high performance in detection tasks, an important objective is the identification of features that can assist with AD diagnosis in clinical settings.   Some studies highlighted various semantic and lexico-syntactic features, such as the proportion of personal pronouns and average sentence length \cite{Balagopalan}.

Beyond the ADReSS Challenge, researchers have also explored the Pitt Corpus \cite{Becker1994} within the Dementia Databank. Some studies have constructed models from scratch, introducing various model variations \cite{karlekar-etal-2018-detecting}, while others have leveraged pre-existing models \cite{Matosevic}. Among these, some studies have solely used text transcripts \cite{Guo2020}, while others have focused exclusively on audio recordings \cite{Guo2021}. Only a few have integrated multiple modalities, incorporating both audio and text data for dementia detection \cite{Ilias_2022, sarawgi2020multimodal}.

In existing literature, it has been suggested that a multimodal approach that integrates different types of data—such as audio, text, and time-stamps—can potentially offer a more comprehensive and effective framework for the classification of dementia. Traditional methods often rely on a single data type, which may not capture the complexity of the condition. The utilization of diverse data sources could provide a more nuanced understanding, thereby improving diagnostic accuracy. Furthermore, data augmentation techniques have been explored as a way to make the diagnostic models more robust and efficient, and their impact on dementia classification can be further examined.

\section*{Methodology}
We evaluate two data modalities, audio and text, as well as text-based synonym data augmentation and the inclusion of explicit timestamps as a model input. 

\subsection*{Audio model}
\begin{figure}[htbp]
  \caption{Architecture of Audio Model}
  \includegraphics[width=\linewidth]{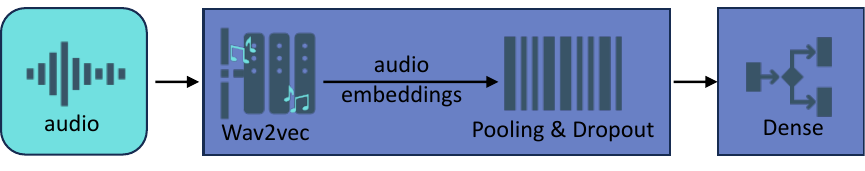}
  \label{fig:audio}
\end{figure}

We created an audio model (Figure \ref{fig:audio}) which was fine-tuned using Wav2vec as the baseline representation. The audio data was processed through Wav2vec to obtain audio embeddings, which were then subjected to a dense layer for binary classification, using binary cross-entropy loss as the evaluation metric. Note that the weights from the pretrained Wav2vec was frozen during the training and only the other layers of architecture were updated.

\subsubsection*{Wav2vec}
Wav2vec \cite{Baevski} is a self-supervised convolutional architecture that transforms audio waveforms into representative embeddings. Initially trained on unlabeled audio data, these embeddings are then processed through a transformer for a masked task. In this task, half of the audio embeddings are masked and predicted by the remaining unmasked portions. Wav2vec is particularly useful in speech recognition tasks due to its adaptability to various audio recordings and its inherently robust performance.

\subsection*{Text model}
\begin{figure}[htbp]
  \caption{Architecture of Text Model}
  \includegraphics[width=\linewidth]{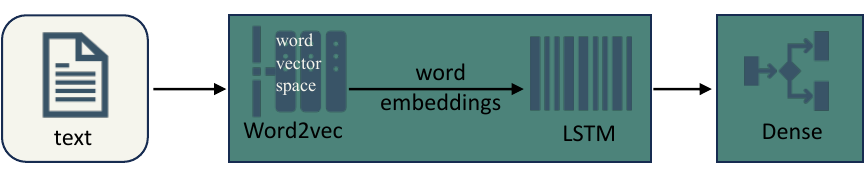}
  \label{fig:text}
\end{figure}

The text model (Figure \ref{fig:text}) incorporates the embedding layers from Word2vec and utilizes an LSTM model connected to a dense layer for classification tasks.

\subsubsection*{Word2vec}
Word2vec \cite{mikolov2013efficient} is a feed-forward neural network designed to produce vector representations of words. It uses surrounding words as input to generate these vectors, capturing semantic relationships between the words. The resulting vectors effectively position semantically similar words closer in the vector space. Again, similar to the audio model, the weights from the pretrained Word2vec was frozen during the training and only the other layers of architecture were updated. 

\subsubsection*{LSTM}
 We employed an LSTM model with 16 units to process embedded sentences and used a dropout and recurrent dropout rate of 0.2. A dense layer with sigmoid activation is appended to the LSTM layer to perform binary classification.

\subsection*{Timestamps} 
\begin{figure}[!h]
  \begin{minipage}[c]{0.5\textwidth}  
  \centering
  \includegraphics[width=\linewidth]{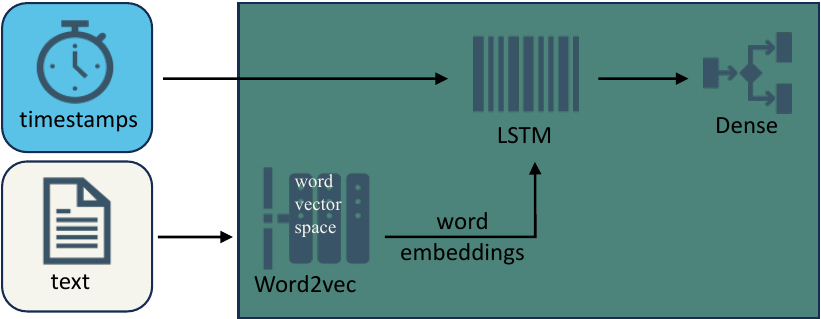}
  \caption{Architecture of Text+Timestamps Model}
  \label{fig:text+time}
  \end{minipage} 
  \begin{minipage}[c]{0.5\textwidth} 
  \centering
  \includegraphics[width=\linewidth]{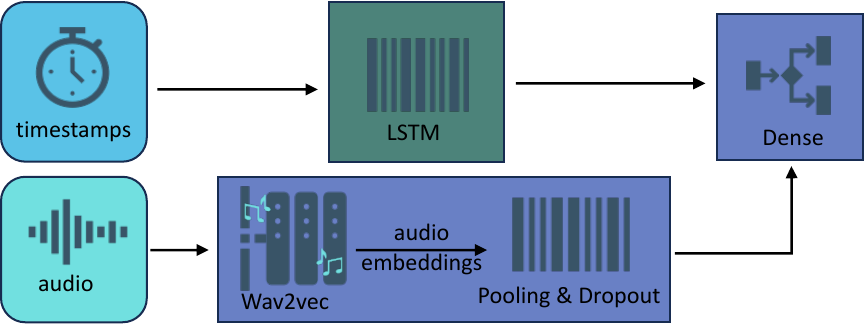}
  \caption{Architecture of Audio+Timestamps Model}
  \label{fig:audio+time}
  \end{minipage} 
\end{figure}

Timestamps for each word were extracted from the corpus. In models that combined text and time (Figure \ref{fig:text+time}), these timestamps were concatenated with the word embeddings before feeding them as input into subsequent layers. In models incorporating audio and time (Figure \ref{fig:audio+time}), timestamps were processed through an LSTM layer, following the extraction of audio embeddings, which were then passed through an average pooling layer and a dropout layer prior to classification.

\subsection*{Concatenated model} 
\begin{figure}[!h]
  \begin{minipage}[c]{0.5\textwidth}  
  \centering
  \includegraphics[width=\linewidth]{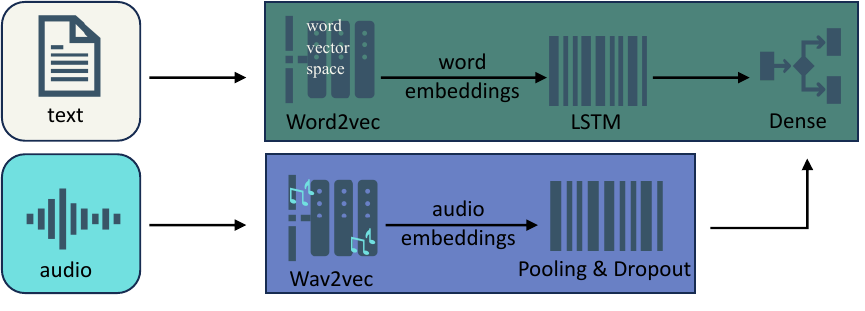}
  \caption{Architecture of Audio+Text Model}
   \label{fig:audio+text}
 \end{minipage} 
  \begin{minipage}[c]{0.5\textwidth}  
  \centering
  \includegraphics[width=\linewidth]{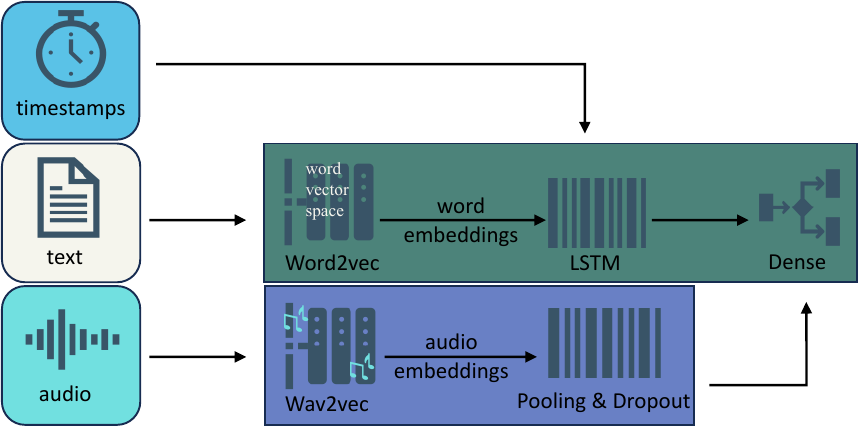}
  \caption{Architecture of Audio+Text+Timestamps Model}
  \label{fig:audio+text+time} 
  \end{minipage} 
\end{figure}

In the concatenated audio-text model (Figure \ref{fig:audio+text}), word embeddings from the text model were processed through an LSTM layer. The audio model was then passed through the same average pooling and dropout layers before the concatenation with the text model. A final dense layer was added for classification tasks. We also developed a model combining audio, text, and timestamps (Figure \ref{fig:audio+text+time}). The architecture for text and timestamps remained consistent with their individual models, as did the audio model. These were then concatenated for the final classification task.

\subsection*{Data augmentation} 
Due to the lack of a substantial volume of data points in the original dataset, we implemented data augmentation techniques. Specifically, we employed the Synonym Replacement (SR) method \cite{wei-zou-2019-eda} where a synonym for a word is identified and used to create a duplicated sentence with the original word replaced by its synonym. Each word was replaced by its synonym twice (n=2). For instance, if a sentence comprises 5 words, all of which have synonyms available in the NLTK dictionary, five new sentences will be generated, each having one original word replaced by a synonym. This is in addition to the original sentence and, therefore, can largely expand the data set. 

\section*{Experiments}
\label{sec:experiments}

\subsection*{Datasets and Data Preprocessing}
\textbf{Dataset source} Our dataset comes from the ``Pitt Cookie Theft'' folder of the Dementia Databank  \cite{Becker1994} (\url{https://dementia.talkbank.org}). This dataset contains participants' responses when asked to describe what they see in stimulus photographs. Unique to this dataset in the Databank is the inclusion of timestamps for each word, which enabled us to use it for our multimodal models.

\textbf{Data Preparation} Both audio and text data were divided into individual sentences, with each sentence serving as a data point for training. The dataset contains 9,447 data points, of which 3,873 are dementia-related and 5,574 are control points. These numbers were derived by categorizing the sentences spoken by investigators as control, as well as those spoken by patients in the control dataset from the Cookie folder. Dementia datapoints, conversely, only include sentences spoken by patients classified as dementia cases in the same folder.

For audio preprocessing, the dataset was first processed through a Wav2vec feature extractor, ensuring compatibility with the sampling rates used during the model's pre-training. The text data underwent tokenization using a custom dictionary, mapping words to their corresponding pre-trained word2vec embeddings in Gensim's `word2vec-google-news-300' \cite{rehurek2011gensim}. Words without corresponding embeddings were marked as Out-of-Vocabulary (OOV) and were represented by zero vectors.

Timestamps for each word were retained, indicating the starting and ending times. The timestamp for the first word in each sentence was normalized to begin at 0 and processed as decimal digits.

Four different dataset conditions were created:\\
\textbf{Original Condition:} The original dataset with 9,447 datapoints, including 3,873 dementia and 5,574 control datapoints.\\
\textbf{Shorts-Removed Condition:} Excluded sentences shorter than two words, resulting in 4,318 control and 3,368 dementia datapoints.\\
\textbf{Original-Augmented Condition:} Augmented from the dataset in Original Condition, leading to 31,273 control and 22,664 dementia datapoints.\\
\textbf{Shorts-Augmented Condition:} Augmented from the dataaset in Short-Removed Condition, yielding 28,964 control and 22,039 dementia datapoints.

For all conditions, the datasets were divided into training and test sets using a 4:1 ratio. Furthermore, the training datasets were also split into training and validation sets with a 4:1 ratio, facilitating 5-fold cross-validation for hyperparameter optimization.

\subsection*{Experimental Setup} 
All models were trained for 50 epochs with a batch size of 16. The objective was to minimize binary cross-entropy loss. To prevent overfitting, early stopping was implemented, which halted training if the validation loss failed to decrease for 10 consecutive epochs. The implementation was carried out using the TensorFlow Keras library \cite{chollet2015keras}.

\subsection*{Results} The experiment was conducted with five separate and independent train-validation splits to ensure generalizability and reliability. The results were then averaged, and both the mean and standard deviation were reported. Evaluation metrics included: accuracy, precision, recall, F1 score, and AUC ROC scores. The highest validation scores for each metric are reported in bold.

\begin{table*}[ht]
\centering
  \caption{\label{tab:original}Results from Original Condition}
  \begin{tabular}{|l|l|l|l|l|l|}
  \hline
  \bfseries Original-augmented & \bfseries Accuracy & \bfseries Precision & \bfseries Recall & \bfseries F1-score & \bfseries AUROC \\
  \hline
  Audio & 0.6484±0.008 & 0.593±0.019 & 0.4425±0.063 & 0.5039±0.032 & 0.7085±0.011\\
  \hline
  Text & \textbf{0.691±0.034} & 0.6484±0.07 & 0.6299±0.127 & 0.62765±0.027 & 0.7638±0.046\\
  \hline
  Audio+Time & 0.6123±0.006 & 0.5565±0.022 & 0.3286±0.039 & 0.4115±0.026 & 0.6517±0.011\\
  \hline
  Text+Time & 0.6909±0.013 & \textbf{0.6537±0.036} & 0.5566±0.4463 & 0.5995±0.022 & \textbf{0.7647±0.018}\\
  \hline
  Audio+Text & 0.6731±0.04 & 0.5958±0.047 & \textbf{0.6852±0.097} & \textbf{0.6341±0.048} & 0.7448±0.045\\
  \hline
  Audio+Text+Time & 0.6539±0.031 & 0.5874±0.07	& 0.5301±0.138 & 0.55±0.087 & 0.7161±0.043\\
  \hline
  \end{tabular}
\end{table*}

\begin{table*}[ht]
\centering
  \caption{\label{tab:original-augmented}Results from Original-augmented Condition}
  \begin{tabular}{|l|l|l|l|l|l|}
  \hline
  \bfseries Original-augmented & \bfseries Accuracy & \bfseries Precision & \bfseries Recall & \bfseries F1-score & \bfseries AUROC \\
  Audio & 0.6038±0.036 & 0.5846±0.021 & 0.1831±0.04 & 0.2764±0.044 & 0.6336±0.003\\
    \hline
  Text & 0.8294±0.005 & \textbf{0.8339±0.022} & 0.751±0.042 & 0.7892±0.013 & 0.9208±0.002\\
    \hline
  Audio+Time & 0.6306±0.068 & 0.5994±0.012 & 0.3673±0.044 & 0.4542±0.033 & 0.6759±0.008\\
    \hline
  Text+Time & \textbf{0.8344±0.006} & 0.8267±0.035 & 0.7721±0.043 & \textbf{0.797±0.009} & \textbf{0.9236±0.005}\\
    \hline
  Audio+Text & 0.825±0.013 & 0.7978±0.039 & \textbf{0.7859±0.056} & 0.7899±0.019 & 0.9124±0.015\\
    \hline
  Audio+Text+Time & 0.8315±0.014 & 0.8212±0.034 & 0.767±0.012 &	0.7927±0.014 & 0.9177±0.014\\
    \hline
  \end{tabular}
\end{table*}

\begin{table*}[ht]
\centering  
\caption{\label{tab:shorts-removed}Results from Shorts-removed Condition}
  \begin{tabular}{|l|l|l|l|l|l|}
  \hline
  \bfseries Shorts-removed & \bfseries Accuracy & \bfseries Precision & \bfseries Recall & \bfseries F1-score & \bfseries AUROC \\
  \hline
  Audio & 0.5958±0.014 & 0.587±0.046 & 0.2945±0.108 & 0.3811±0.084 & 0.624±0.022\\
  \hline
  Text & \textbf{0.6861±0.027} & \textbf{0.623±0.042} & 0.6951±0.058 & \textbf{0.6545±0.022} & \textbf{0.7593±0.024}\\
  \hline
  Audio+Time & 0.5897±0.009 & 0.5631±0.034 & 0.3554±0.07 & 0.4306±0.054 & 0.6163±0.01\\
  \hline
  Text+Time & 0.6683±0.029& 0.6197±0.043 & \textbf{0.7011±0.137} & 0.6494±0.032 & 0.7353±0.047\\
  \hline
  Audio+Text & 0.6052±0.014 & 0.5669±0.024 & 0.5255±0.152 & 0.534±0.082 & 0.6482±0.014\\
  \hline
  Audio+Text+Time & 0.6257±0.065 & 0.5724±0.075 & 0.5239±0.151 & 0.5415±0.107	& 0.6769±0.081\\
  \hline
  \end{tabular}
\end{table*}

\begin{table*}[ht]
\centering  
  \caption{\label{tab:shorts-augmented}Results from Shorts-augmented Condition} 
  \begin{tabular}{|l|l|l|l|l|l|}
  \hline
  \bfseries Shorts-augmented & \bfseries Accuracy & \bfseries Precision & \bfseries Recall & \bfseries F1-score & \bfseries AUROC \\
  \hline
  Audio & 0.5954±0.001 & 0.5989±0.025 & 0.1939±0.029 & 0.2914±0.03 & 0.6258±0.008\\
  \hline
  Text & 0.841±0.01 & 0.8212±0.014 & \textbf{0.8089±0.024} & 0.8148±0.014 & 0.9276±0.009\\
  \hline
  Audio+Time & 0.6254±0.005 & 0.5894±0.023 & 0.4298±0.05 & 0.4951±0.032 & 0.6692±0.008\\
  \hline
  Text+Time & \textbf{0.8478±0.003} & \textbf{0.8375±0.02} & 0.8039±0.207 & \textbf{0.8199±0.007} & \textbf{0.9345±0.005}\\
  \hline
  Audio+Text & 0.835±0.02 & 0.8154±0.036 & 0.7982±0.039 & 0.80591±0.023 & 0.9216±0.02\\
  \hline
  Audio+Text+Time & 0.8451±0.003 &	0.83646±0.027 &	0.7992±0.028 &	0.8166±0.04 & 0.931±0.005\\
  \hline
  \end{tabular}
\end{table*}

Our results highlight the challenges and opportunities associated with utilizing multimodal data for dementia detection. Specifically, as evidenced in Table \ref{tab:original}, Figures 7a and 8a, unimodal audio models underperformed compared to their text counterparts. The audio+time model as in Figures 7a and 8d also yielded suboptimal results, further highlighting that audio, as a modality, may be challenging to engineer a useful feature representation for, with current state-of-the-art methods such as Wav2vec not providing enough specificity and expressiveness of the relevant features. On the other hand, the text model (Table \ref{tab:original}, Figures 7a and 8b) excelled across the board, and its efficacy was only augmented when coupled with time, as demonstrated by the superior performance of the text+time model (Table \ref{tab:original}, Figures 7a, 8e and 8f). 

We observed higher standard deviations in some modalities, primarily in the audio-based models. This suggests that the model was more prone to poor fitting in several data splits.

As observed in Table \ref{tab:shorts-removed} and Figures 7c\footnote{The audio+text+time model we saved had an above-average performance}, the exclusion of shorter sentences during preprocessing did not bring significant improvement in the overall model performance. This indicates that the initial preprocessing strategy was robust and captured the most important aspects of the data for the task at hand.

Table \ref{tab:original-augmented}, Figures 7b and Table \ref{tab:shorts-augmented}, Figure 7d illustrate the significant improvement achieved as a result of our data augmentation techniques. AUROC scores in models incorporating text data surpassed 90\% Figure 8b, 8c, 8d and 8e), and both accuracy and F1 scores were consistently above 80\%. This uplift in performance metrics strongly suggests that the augmented data captured richer semantic and syntactic features essential for dementia detection.

The relative consistency in high performance across metrics for text-based models further emphasizes the influence that the textual modality has on the overall multimodal models. Even when coupled with lower-performing audio-based models, the text models lifted the combined model's performance to a more satisfactory level.

\begin{figure*}[!htb]
  \centering
    \begin{minipage}{.5\textwidth}
    \centering
    \includegraphics[width=.8\linewidth]{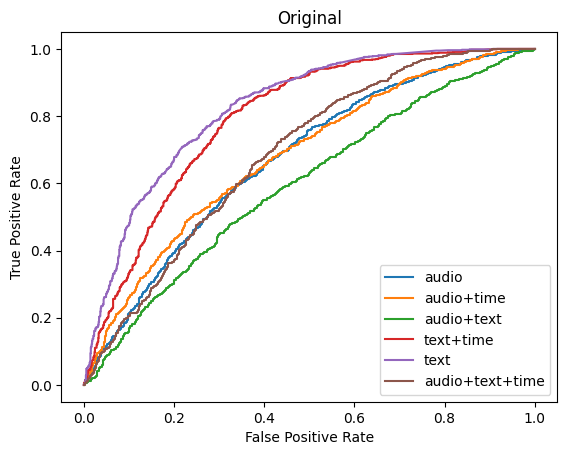}
    \caption*{(a) Original condition}
    \label{fig:sub1}
  \end{minipage}%
  \begin{minipage}{.5\textwidth}
    \centering
    \includegraphics[width=.8\linewidth]{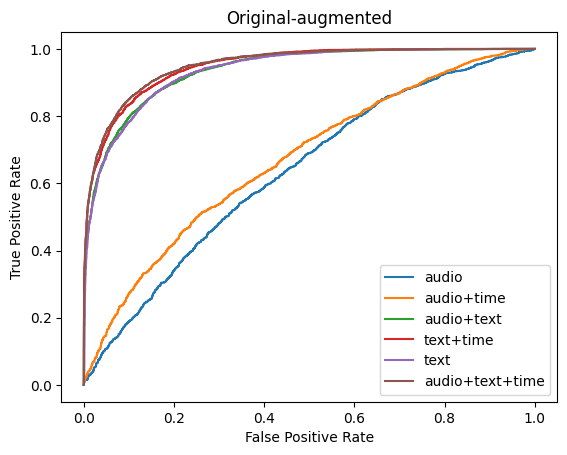}
    \caption*{(b) Original-augmented condition}
    \label{fig:sub2}
  \end{minipage}
  \\ 
  \begin{minipage}{.5\textwidth}
    \centering
    \includegraphics[width=.8\linewidth]{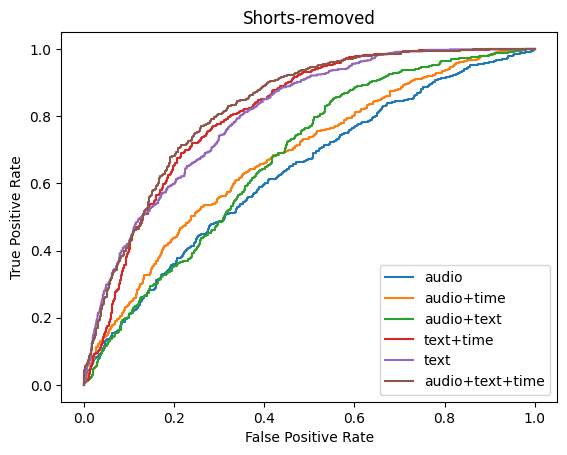}
    \caption*{(c) Shorts-removed condition}
    \label{fig:sub3}
  \end{minipage}%
  \begin{minipage}{.5\textwidth}
    \centering
    \includegraphics[width=.8\linewidth]{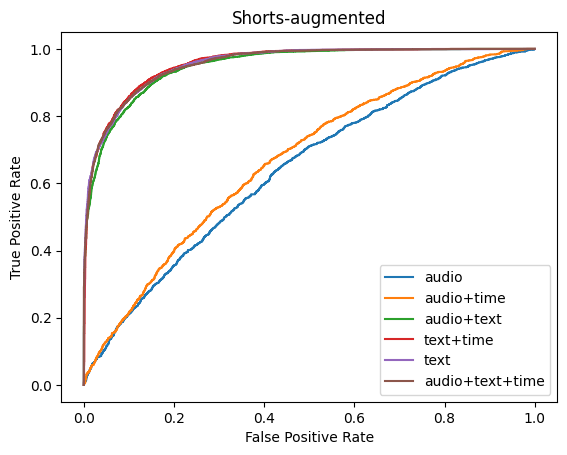}
    \caption*{(d) Shorts-augmented condition}
    \label{fig:sub4}
  \end{minipage}
    \caption{AUC-ROC curves in four conditions}
    \label{fig:figure7}
\end{figure*}

\begin{figure*}[!htb]
    \begin{minipage}{.33\textwidth}
    \centering
    \includegraphics[width=.8\linewidth]{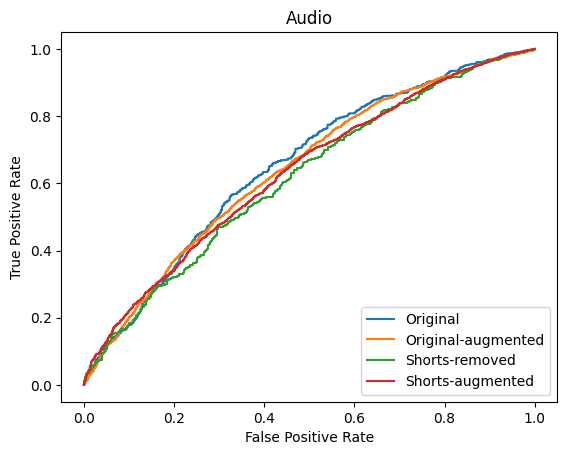}
    \caption*{(a) Audio}
    \label{fig:sub5}
  \end{minipage}%
  \begin{minipage}{.33\textwidth}
    \centering
    \includegraphics[width=.8\linewidth]{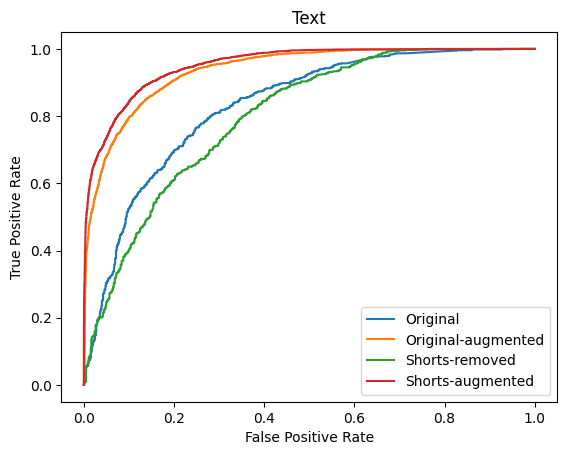}
    \caption*{(b) Text}
    \label{fig:sub6}
  \end{minipage}%
  \begin{minipage}{.33\textwidth}
    \centering
    \includegraphics[width=.8\linewidth]{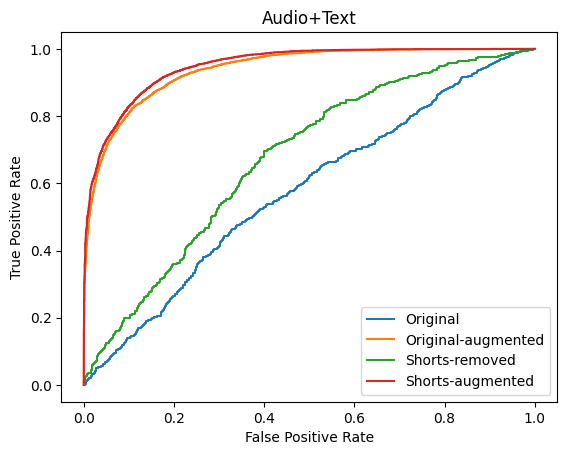}
    \caption*{(c) Audio+Text}
    \label{fig:sub7}
  \end{minipage}
  \\ 
  \begin{minipage}{.33\textwidth}
    \centering
    \includegraphics[width=.8\linewidth]{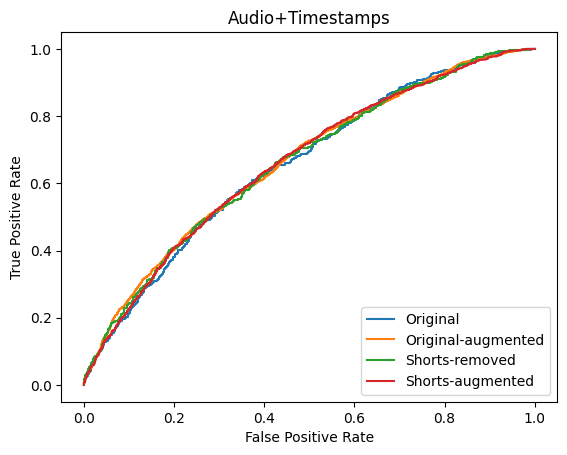}
    \caption*{(d) Audio+Timestamps}
    \label{fig:sub8}
  \end{minipage}%
  \begin{minipage}{.33\textwidth}
    \centering
    \includegraphics[width=.8\linewidth]{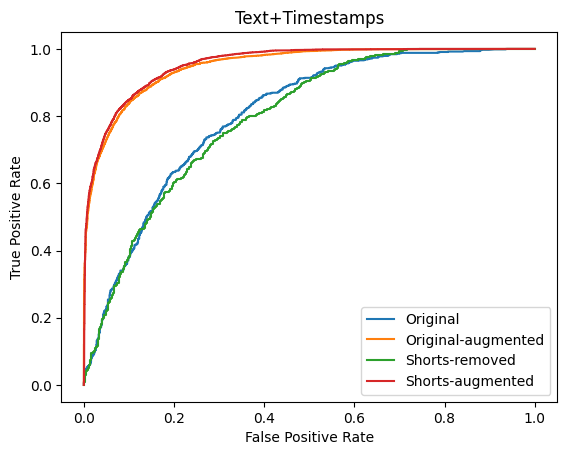}
    \caption*{(e) Text+Timestamps}
    \label{fig:sub9}
  \end{minipage}%
  \begin{minipage}{.33\textwidth}
    \centering
    \includegraphics[width=.8\linewidth]{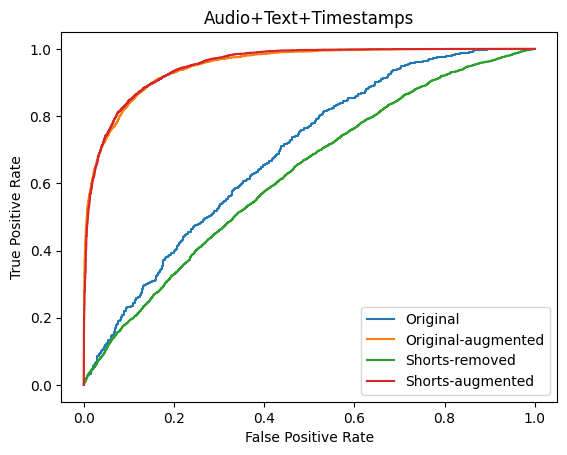}
    \caption*{(f) Audio+Text+Timestamps}
    \label{fig:sub10}
  \end{minipage}%
    \caption{AUC-ROC curves of six types of models}
    \label{fig:figure8}
\end{figure*}

\subsection*{Qualitative Error Analysis}
We conducted a qualitative error analysis to understand which types of sentence archetypes were frequently misclassified. We conducted this error analysis for (1) false positives vs. false negatives, (2) augmented vs. non-augmented sentences, and (3) short sentences removed vs. short sentences augmented.

\textbf{False Positives:} Our text model tended to misclassify certain types of sentences from control patients as originating from dementia patients. These sentences generally exhibited one or more of the following characteristics:
\begin{itemize}
  \item {\textbf{Noun-Phrase Sentences}: Examples include 'curtain on the window,' 'down on this side of the picture.'}
  \item {\textbf{Ungrammatical Sentences}: Sentence types that are uttered by patients in the control group but are slightly unnatural. Examples include 'the boy is uh taking cookies out of the cookie jar,' and 'uh mother’s drying dishes,' ‘that’s real good then.’}
  \item {\textbf{Repetition}: The repetition of patients’ sentences from the investigator, ‘climbing a stool.'}
\end{itemize}

\textbf{False Negatives:} Sentences from dementia patients that were misclassified as from control groups include:
\begin{itemize}
  \item {\textbf{Correct and Transcribed:}Sentences that were grammatically correct and transcribed correctly. Examples include 'that’s about all,' and 'and the girl.'}
  \item {\textbf{Short and Correct:}Examples include sentences like 'here,' ‘okay.’}
  \item {\textbf{Common responses:}Sentences that patients often respond to or ask and are transcribed correctly 'okay,' 'that’s terrible,' 'that’s about it, right?'}
\end{itemize}

In the original-augmented model, the misclassified sentences are as follows:
\begin{itemize}
  \item {\textbf{Unlikely connotations:} Augmented sentences sometimes led to unlikely or misleading connotations, affecting the model's prediction accuracy.}
  \begin{itemize}
  \item[\checkmark] {‘I’ve got the tape recorder on so.’ \\
    (original, control, predicted as control)}
  \item[\checkmark] {‘I’ve got the videotape recorder on so.’ \\
    (augmented, control, predicted as dementia)}
   \item[\checkmark]{I’ve got the tape registrar on so.’ \\
    (augmented, control, predicted as dementia)}
   \end{itemize} 
  \item {\textbf{Word Usage:} Augmented words common to control data were sometimes present in sentences from dementia patients, leading to misclassification.}
  \begin{itemize}
  \item[\checkmark]{‘It shows the mother in the kitchen wiping dishes.’ \\
        (original, dementia, predicted as dementia)}
  \item[\checkmark]{‘It \textbf{testify} the mother in the kitchen wiping dishes.’\\
        (augmented, dementia, predicted as control)}
  \end{itemize}       
 \item  {\textbf{Augmented Correct Sentences:} Sentences that were originally grammatical but became ungrammatical and misclassified after augmentation. For example: }
 \begin{itemize}
 \item[\checkmark]{‘The little girl’s standing there.’\\
        (original, dementia, predicted as dementia)}
 \item[\checkmark]{‘The little miss standing there.’ \\
       (augmented, dementia, predicted as control)}
  \end{itemize}      
\end{itemize}
We also analyzed the model's performance under conditions where short sentences were either removed or augmented. The Shorts-removed condition revealed that the nature of incorrectly predicted sentences generally parallels that of the original dataset, minus the influence of short sentences. This suggests that the presence or absence of short sentences in the data does not dramatically alter the types of errors the model makes, which implies that the model's predictive capabilities are not significantly affected by sentence length alone. Interestingly, the errors made by the model in the Shorts-augmented condition closely resembled those in the original-augmented condition. This might point toward the robustness of the augmented dataset's influence on model behavior, regardless of the presence or absence of short sentences. This further emphasizes that while data augmentation significantly enhances the model's overall performance, it does not necessarily change the nature of the mistakes made by the model in prediction.

\section*{Discussion and Conclusion}
\label{sec:conclusion}

We have presented a novel approach to dementia detection by leveraging multimodal data consisting of audio and text. Utilizing pre-trained models like Wav2vec and Word2vec, we have demonstrated robust performance when the model is sufficiently trained on a large dataset. Importantly, the presence of text data seems to bolster the performance of the model significantly, even compensating for lower-performing audio data. This suggests that text-based data can be a critical component in improving the diagnostic accuracy of dementia detection systems.

The study also highlights the limitations of individual modalities and the importance of integrating multiple types of data for more accurate results. While the contribution of audio and timestamp data was relatively modest, their inclusion in a multimodal framework did sometimes lead to marginal improvements, however slight. Further work is required to identify whether there are more successful ways to incorporate both text and audio data into a single model.

Our results suggest that Wav2vec audio representations are insufficient for significantly bolstering model performance. This result is surprising in light of previous work which was able to classify autism using audio from naturalistic home videos using Wav2vec representations \cite{chi2022classifying}. Part of the success of these prior efforts are likely due to the relatively structured nature of the input audio, where the structure was imposed by the mechanics of a mobile game \cite{kalantarian2019guess, kalantarian2018gamified, washington2022improved, kalantarian2018mobile, kalantarian2019labeling}. Future work is therefore required to properly extract audio features which are relevant to the classification of dementia.

\bibliography{sample}



\section*{Availability of materials and data}
The datasets generated during and/or analysed during the current study are available from the corresponding author on reasonable request.

\section*{Code availability}
Code available at \url{https://anonymous.4open.science/r/dementia-176E}.

\end{document}